\documentclass[prl,twocolumn,superscriptaddress,showpacs,nofootinbib,reprint,floatfix,aps]{revtex4-1}

\pdfoutput=1
\usepackage{graphicx,amsmath,amssymb,bm}
\graphicspath{{plots/}} 

\newcommand{\Trel}{T_{\rm rel}}
\newcommand{\beqn}{\begin{equation}}
\newcommand{\eeqn}{\end{equation}}
\newcommand{\bea}{\begin{eqnarray}}
\newcommand{\eea}{\end{eqnarray}}
\newcommand{\ba}{\begin{align}}
\newcommand{\ea}{\end{align}}

\newcommand{\fmi}{\, \text{fm}^{-1}}

\newcommand{\vlowk}{V_{{\rm low}\,k}}

\begin{document}
\title{Neutron matter based on consistently evolved chiral three-nucleon interactions}

\author{K.\ Hebeler}
\email[E-mail:~]{hebeler.4@osu.edu}
\affiliation{Department of Physics, The Ohio State University, Columbus, OH 43210, USA}

\author{R.\ J.\ Furnstahl}
\email[E-mail:~]{furnstahl.1@osu.edu}
\affiliation{Department of Physics, The Ohio State University, Columbus, OH 43210, USA}

\begin{abstract}
We present the first results for the neutron matter equation of state (EOS) using nucleon-nucleon and 
three-nucleon chiral effective field theory interactions that are consistently evolved 
in the framework of the Similarity Renormalization Group (SRG). 
The dependence of the EOS on the SRG resolution scale is greatly
reduced when induced three-nucleon forces (3NF) are included and the residual variation,
which in part is from missing induced four-body interactions,
is comparable to estimated many-body perturbation theory truncation errors. 
The relative growth with decreasing resolution of the 3NF contributions 
to the energy per neutron is of natural size, but it accelerates at
the lowest resolutions where strong renormalization of the long-range
3NF matrix elements is also observed. 
\end{abstract}


\maketitle

Chiral effective field theory (EFT)~\cite{Epelbaum_RevModPhys} offers a systematic expansion of 
nuclear forces well suited to meet the calculational challenges of neutron matter, which 
span the extremes of low-density universal properties to the dense
matter in neutron stars. Conversely, neutron matter provides a powerful laboratory for testing chiral EFT 
power counting at relevant nuclear densities, since only long-range three-nucleon forces (3NF) contribute at 
next-to-next-to-leading order (N$^2$LO)~\cite{Hebeler_PNM} and there are no new parameters for three-nucleon (3N)
and four-nucleon interactions at next-to-next-to-next-to-leading order (N$^3$LO)~\cite{Epelbaum_RevModPhys}. For example, it 
was recently shown that 3NF at N$^3$LO give relatively large contributions to the neutron matter equation of 
state (EOS)~\cite{Tews_N3LO, Krueger_N3LO}, which may indicate that a chiral EFT with explicit 
delta degrees of freedom would be more efficient.

\begin{figure}[b!]
\includegraphics[scale=0.45,clip=]{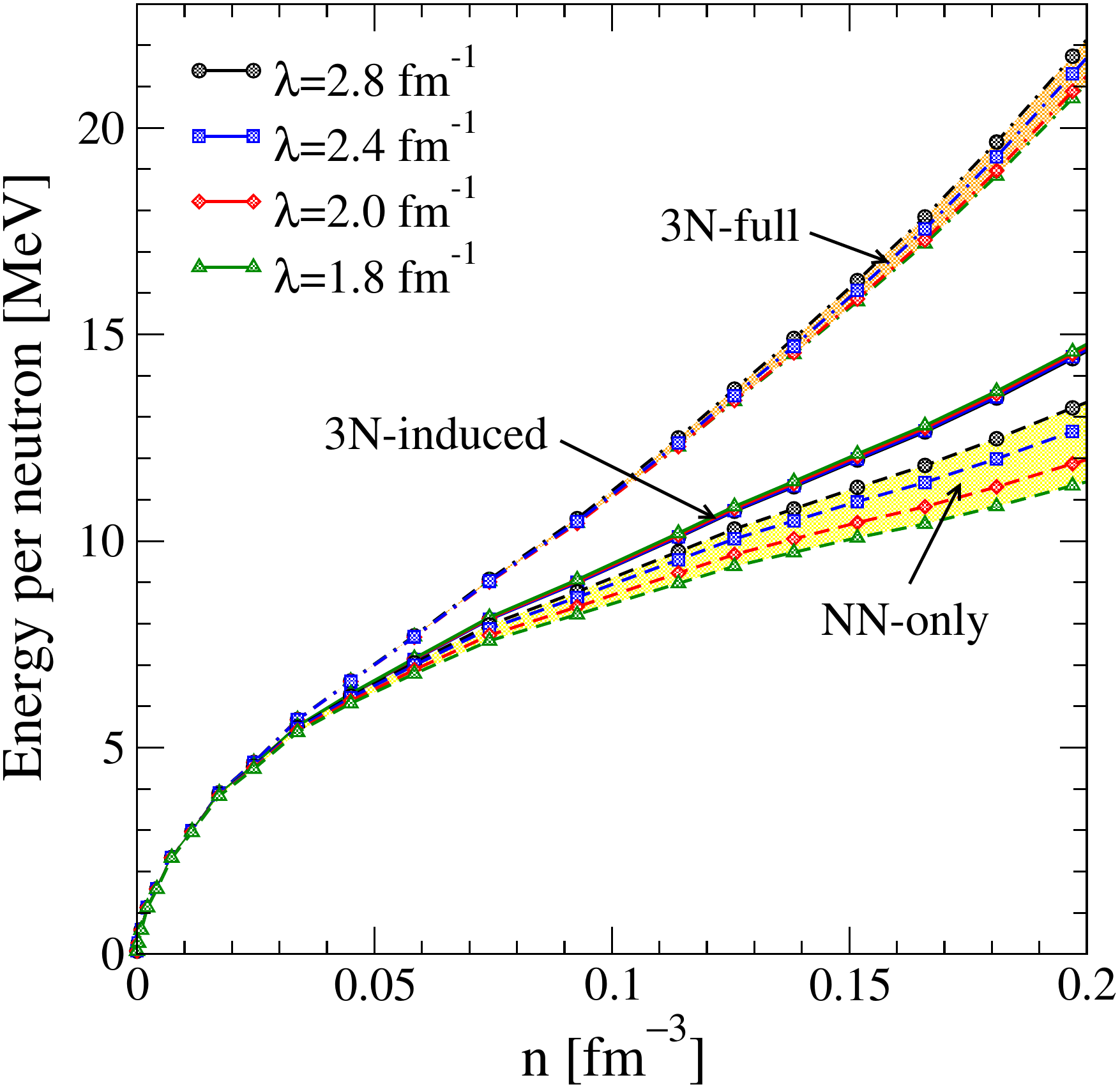}
\caption{(Color online) Energy per neutron as a function of neutron density for different SRG resolution scales.
The results are grouped according to whether no induced 3NF are included
(NN-only), the induced 3NF are included but no initial 3NF (3N-induced),
or initial and induced 3NF are included (3N-full). 
The initial interaction is the 500\,MeV N$^3$LO NN potential from
Ref.~\cite{N3LO} combined with the N$^2$LO 3NF using the consistent low-energy constants
$c_1 = - 0.81\,\rm{GeV}^{-1}$ and $c_3 = - 3.2\,\rm{GeV}^{-1}$~\cite{Hebeler_PNM}.}
\label{fig:Erho}

\end{figure}

At present, the largest uncertainties in microscopic calculations of 
neutron matter based on chiral EFT interactions are because the 
low energy constants in the Hamiltonian are not all well determined.
This leads to uncertainties in observables such as the nuclear
symmetry energy and radii of neutron stars~\cite{Hebeler_NS}. 
However, direct calculations based on chiral interactions using many-body perturbation theory (MBPT) 
also have non-negligible theoretical uncertainties due to truncations of the many-body 
expansion~\cite{Tews_N3LO, Krueger_N3LO}. Renormalization group (RG) evolution of nuclear 
interactions to lower resolution scales significantly improves the 
convergence of MBPT, 
but in prior calculations three-nucleon interactions have not been 
evolved consistently~\cite{Hebeler_PNM}. Here we present the first results for 
the neutron matter equation of state based on 
consistently evolved chiral nucleon-nucleon (NN) and 3N forces (see Fig.~\ref{fig:Erho}). These 
results show how RG transformations can enable simplified and efficient many-body
calculations for neutron matter with controlled theoretical error bars.

We build upon Ref.~\cite{Hebeler_3NF_evolution}, which presented a framework
for the simultaneous evolution of nuclear NN and 3N interactions in a
continuous plane-wave (momentum) basis  via the Similarity Renormalization Group (SRG). 
This is an alternative to using a harmonic oscillator basis~\cite{Jurgenson_PRL}.
It provides independent checks of the SRG evolution, easier access to alternative SRG 
generators (see Refs.~\cite{Anderson_BlockSRG,Li_Generators})
and a means to test approximations for induced 3NF~\cite{Hebeler_3NF_evolution}.
The momentum-space matrix elements can be easily
transformed to an oscillator basis for use in calculations
of finite nuclei by configuration interaction~\cite{Jurgenson_PRC}, coupled 
cluster~\cite{Hagen_CC3N, Binder_CC}, in-medium SRG methods~\cite{Tsukiyama_2012, Hergert_2012}, 
self-consistent Gorkov Green's function theory~\cite{Soma_2012zd} or
for nuclear shell-model calculations~\cite{Otsuka}.

\begin{figure*}[t!]
\includegraphics[scale=0.45,clip=]{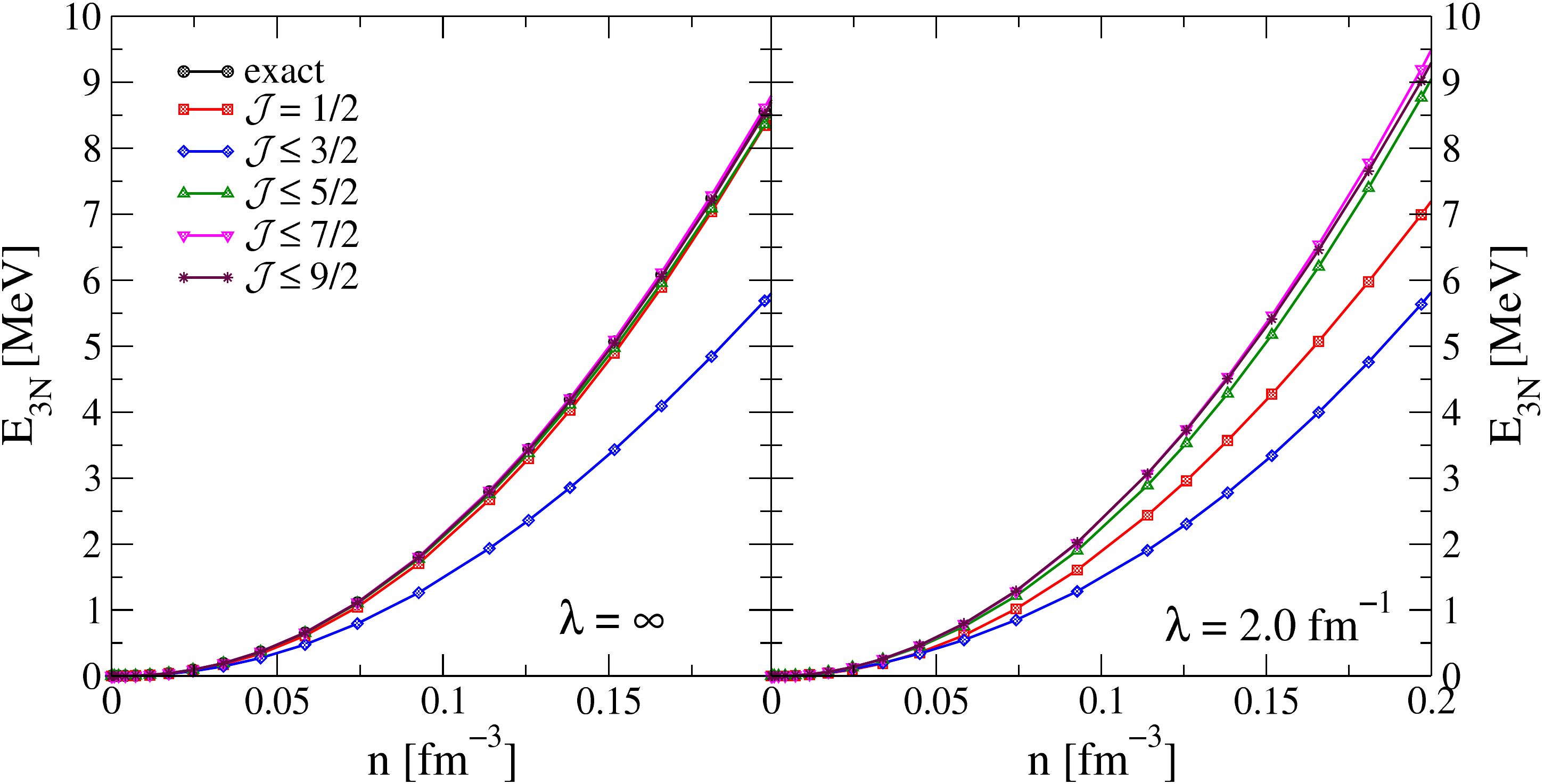}
\caption{(Color online) Partial-wave convergence of the Hartree-Fock 3NF-only
energy per neutron as a function of density at two different resolution scales.
The left panel shows the results before evolution ($\lambda = \infty$) 
and the right panel at $\lambda=2.0\,\fmi$.
The Hamiltonian is the same as for the 3N-full results in Fig.~\ref{fig:Erho}.
For $\lambda=\infty$, only $\mathcal{J} \le 5/2$
give significant contributions so the lines for the higher partial 
waves are nearly indistinguishable from the exact result.}
\label{fig:pw_conv}
\end{figure*}

Most important in the present context is that
these evolved interactions can be directly applied in microscopic calculations of the nuclear EOS.
Applications to the triton in Ref.~\cite{Hebeler_3NF_evolution} verified that nuclear interactions at
low resolution scales contain much weaker couplings between low and high-momentum states.
This renders the nuclear many-body problem more perturbative and therefore more tractable, with all low-energy observables 
preserved exactly when all induced contributions are included. However, recent results for 
finite nuclei using oscillator-evolved 3NF found that high-momentum parts of the chiral two-pion-exchange 
3NF led to significant resolution-scale dependence and overbinding in medium-mass nuclei~\cite{Roth_NCSM, Roth_CC}. 
Neutron matter provides a test laboratory (although limited, because only neutrons) to study if 
these systematics carry over to infinite matter in MBPT with the new evolution scheme.
We have direct access to the form of the 3NF and the scaling behavior of their
contributions to the energy as a function of the resolution scale.

The SRG flow equations we solve can be written in the form~\cite{Bogner_SRG_evolution} 
\beqn
  \frac{d H_s}{ds} = \left[ \eta_s, H_{s} \right] 
  \;,
  \label{dHds}
\eeqn
where $H_s = \Trel + V_s$ denotes the Hamiltonian as a function of the
flow scale parameter $s$, and $\eta_s$ labels the generator of  the RG
transformations. In practice it is more informative to replace $s$ with 
the resolution (or decoupling) scale $\lambda = s^{-1/4}$, which
has units of momentum. Here we choose $\eta_s = [ \Trel, H_s ]$ with the
relative kinetic energy $\Trel$, as in Ref.~\cite{Hebeler_3NF_evolution} and most prior 
investigations. With this $\eta_s$ the flow equation generates a continuous
series of unitary transformations that renormalizes the Hamiltonian (and all
other operators), driving $H_s$ towards a diagonal form in momentum
space~\cite{Bogner_review}. We recast Eq.~\eqref{dHds} into separate flow
equations for the matrix elements of the NN and 3N interactions~\cite{Bogner_SRG_evolution, Hebeler_3NF_evolution} 
and solve them simultaneously in a momentum partial-wave basis.

For the NN forces we use a standard partial-wave basis of the form $\left| p; (LS) J T \right>$, where 
$p$ is the relative momentum and $L$, $S$, $J$ and $T$ are the orbital angular momentum, spin, total angular 
momentum and isospin of the interacting pair. For the three-body basis we choose  
\beqn
 \left| p_i q_i; \left[ (L S) J (l s_i) j \right] 
  \mathcal{J} \mathcal{J}_z (T t_i) \mathcal{T} \mathcal{T}_z \right\rangle
  \equiv 
  \left| p q \alpha \right\rangle_i 
  \;,
\eeqn
where $p_i$ and $q_i$ are the three-body Jacobi momenta of particle $i$. The 
quantum numbers $l$, $s_i=1/2$, $j$ and $t_i=1/2$ are the orbital angular
momentum, spin, total angular momentum and isospin of particle  $i$ relative to
the center-of-mass of the pair with momentum $p$. $\mathcal{J}$ and
$\mathcal{T}$ are  the total three-body angular momentum and isospin quantum
numbers (for details see Refs.~\cite{Gloeckle_book, Stadler}). The 3NF at
chiral order N$^2$LO are independent of the projections $\mathcal{J}_z$ and
$\mathcal{T}_z$~\cite{Epelbaum_RevModPhys}. We use $\alpha$ to abbreviate the angular 
momentum and isospin quantum numbers. 

In the SRG evolution we take all NN interaction matrix elements up through
$J_{\rm{max}} = 7$ into account, and for the 3NF matrix elements we include all
interactions up through $\mathcal{J} = 9/2$  and $J_{\rm{max}} = l_{\rm{max}} =5$. 
In Fig.~\ref{fig:pw_conv} we show the convergence of the partial-wave 3NF 
contributions in the Hartree-Fock approximation as a function of density for the
initial interaction and  after evolution to $\lambda = 2.0 \fmi$. While the
convergence pattern is somewhat altered after evolution, the partial-wave
truncation is reliable at both scales and the results are well converged. 
The exact Hartree-Fock energy can be calculated without a partial wave expansion for 
the unevolved interaction~\cite{Hebeler_PNM,Tolos2008}, and in this case 
the energy through $\mathcal{J} = 9/2$ is converged within 0.4~percent at nuclear saturation density.

The RG evolution of the 3NF can be performed independently for different values of $\mathcal{J}, \mathcal{T}$ and 
three-body parity $\pi_3 = (-1)^{L+l}$. For applications to neutron matter we only need matrix elements for the isospin 
channel $\mathcal{T} = 3/2$. Our calculations are based on antisymmetrized matrix elements of the form
\begin{equation}
  \bigl\langle p q \alpha| \overline{V}_{123} | p' q' \alpha' \bigr\rangle \equiv  \hspace{-1mm} \phantom \langle_i \bigl\langle p q 
  \alpha| \mathcal{A}_{123} V_{123}^{(i)} \mathcal{A}_{123} | p' q' \alpha'   
  \bigr\rangle \hspace{-1.2mm} \phantom \rangle_i
  \;, \label{eq:symm_int}
\end{equation}
where $V_{123}^{(i)}$ is the $i$-th Faddeev component of the three-body 
interaction, $\mathcal{A}_{123} = (1 + P_{123} + P_{132})$ is an
antisymmetrizer, and $P_{123} (P_{132})$ is the cyclic (anti-cyclic)
permutation operator (see Ref.~\cite{Gloeckle_book}). Such antisymmetrized 
matrix elements with two antisymmetrizers, in contrast to partially antisymmetrized 
matrix elements typically used in Faddeev calculations~(see, e.g., Ref~\cite{Skibinski_aPWD}), 
are particularly suitable for solving the SRG flow equations for 3NF (see Ref.~\cite{Hebeler_3NF_evolution}). The
initial matrix elements are generated directly in this form using a novel
automatized partial-wave decomposition for the 3NF~\cite{Golak_aPWD, Skibinski_aPWD}. This 
new method ensures that the interaction is antisymmetrized exactly in spin and isospin space.

Due to the softening of NN and 3N forces during the RG evolution, 
it is possible to apply MBPT for the calculation of the neutron matter equation of state at low 
resolution scales. In the present calculations we resum contributions from 
NN forces in the ladder approximation while contributions from 3NF are calculated in Hartree-Fock approximation.
Based on results of Ref.~\cite{Hebeler_PNM} we expect this approximation to provide the dominant 
3NF contributions at low-resolution scales. We also compute the NN second-order contributions for comparison, which 
allows us to probe the perturbativeness of the NN interactions as a function of the RG scale $\lambda$. The inclusion of higher-order 
diagrams involving 3NF requires significant computational storage because of the coupling of partial waves 
with different $\mathcal{J}$ and $\pi$. Such calculations are currently in progress.

\begin{figure*}[t!]
\includegraphics[scale=0.45,clip=]{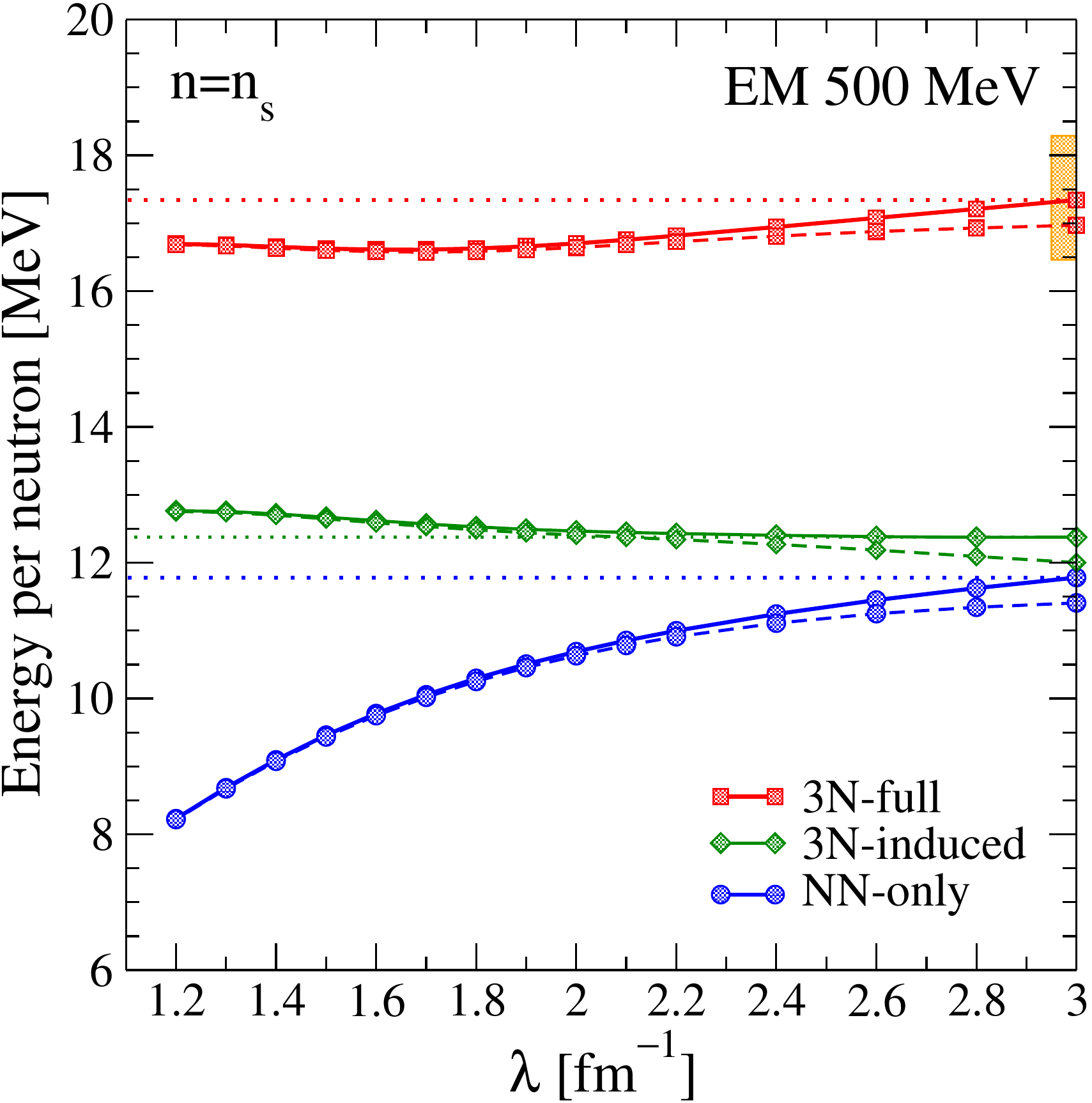}
\hspace{1cm}
\includegraphics[scale=0.45,clip=]{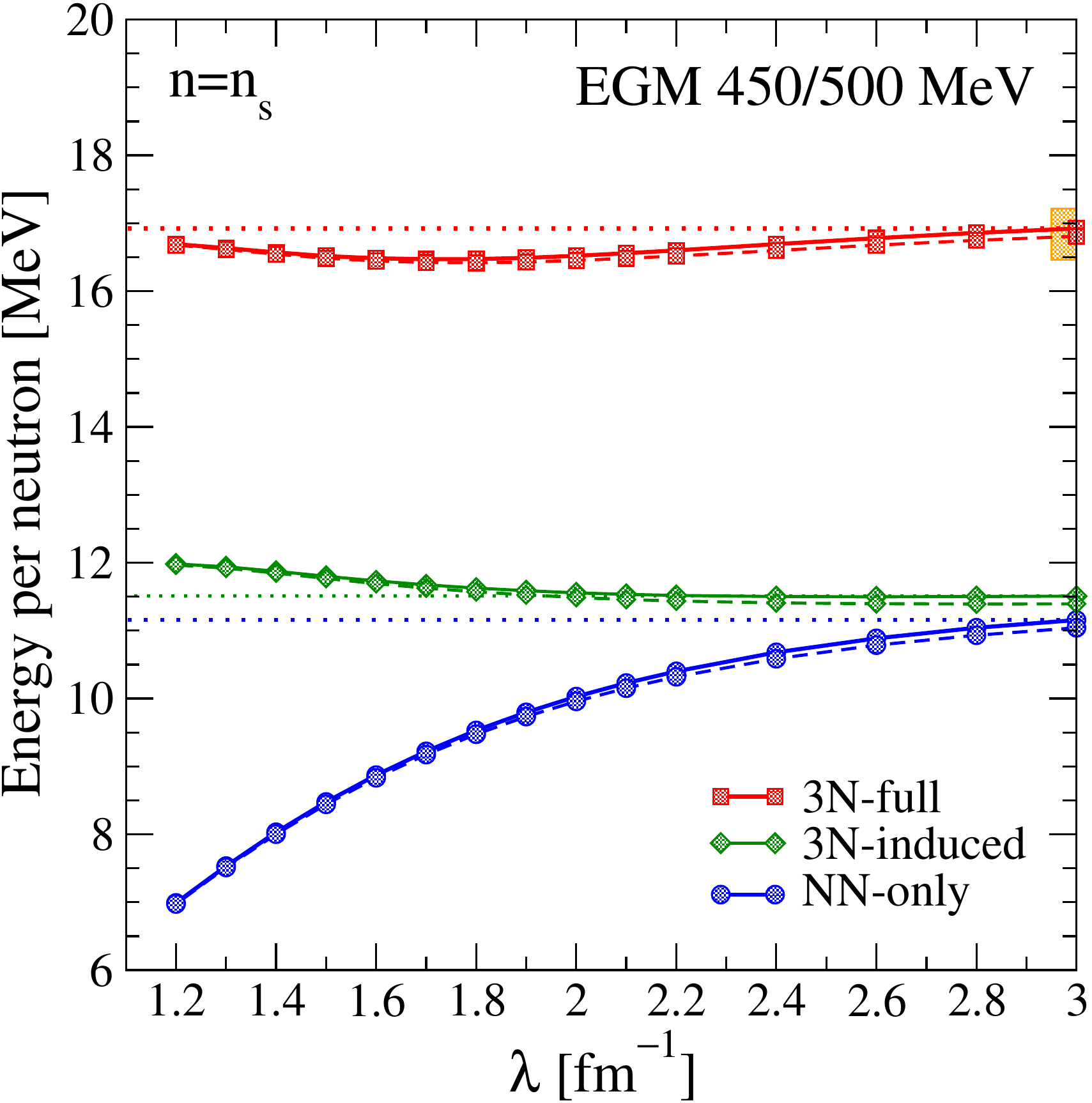}
\caption{(Color online) Energy per neutron as a function of the resolution scale $\lambda$ at nuclear matter 
saturation density $n_s=0.16\,\rm{fm}^{-3}$ for two different interactions. The left panel uses the NN 
and 3N potentials as in Fig.~\ref{fig:Erho} and the right panel uses the N$^2$LO potential ($\Lambda/\tilde{\Lambda} = 450/500$ MeV) 
of Ref.~\cite{Progpart_Epelbaum} plus the consistent 3NF with the low-energy constants $c_1 = - 0.81\,\rm{GeV}^{-1}$ 
and $c_3 = - 3.4\,\rm{GeV}^{-1}$. The same three calculations (NN-only, 3N-induced, and 3N-full) from 
Fig.~\ref{fig:Erho} are used here. The solid lines are energies using a resummed particle ladder sum of the NN 
contributions, with the straight dotted lines marking the energies at $\lambda = 3.0\,\rm{fm}^{-1}$. 
The dashed lines are energies including up to second-order for the NN contributions. The shaded areas at the right 
indicate the magnitude of the attractive second-order contribution from unevolved NN and 3N forces.}
\label{fig:Elambda}
\end{figure*}

In the Hartree-Fock approximation, the 3NF contributions to the energy per volume is given by:
\begin{eqnarray}
 \frac{E_{HF}}{V} &=& \frac{1}{18} \prod_{i=1}^3 \text{Tr}_{\sigma_i} \int \frac{d \mathbf{k}_i}{(2 \pi)^3} \nonumber \\
 && \quad \times \left< 1 2 3 | \mathcal{A}_{123} V_{123} \mathcal{A}_{123} | 1 2 3 \right> n_{\mathbf{k}_1} n_{\mathbf{k}_2} n_{\mathbf{k}_3}
  \;, 
\end{eqnarray}
where $n_{\mathbf{k}}$ are the zero-temperature occupation numbers. Compared to the relations of Refs.~\cite{Hebeler_PNM,Bogner_2005} an extra
factor $1/3$ appears because the 3NF are antisymmetrized in the initial and final states. The spin sum in the Jacobi momentum 
basis can be expressed in the form
\begin{eqnarray}
&& \sum_{S, \mu, \nu} \langle \mathbf{p} \mathbf{q} S \mu \frac{1}{2} \nu | \mathcal{A}_{123} V_{123} \mathcal{A}_{123} | \mathbf{p} \mathbf{q} S \mu \frac{1}{2} \nu \rangle \nonumber \\
&=& \frac{1}{(4 \pi)^2} \sum_{\alpha, \alpha'} \delta_{S S'} \sum_{\bar{L},\mathcal{S}, \mathcal{L}, \mathcal{J}} \hat{\mathcal{S}} \hat{\mathcal{L}} \hat{\mathcal{J}} \sqrt{\hat{J} \hat{j} \hat{J}' \hat{j}' \hat{L} \hat{L}' \hat{l} \hat{l}'} \nonumber \\
&& \times (-1)^{l+l'+\mathcal{L}} \mathcal{C}_{l 0 l' 0}^{\bar{L} 0} \mathcal{C}_{L 0 L' 0}^{\bar{L} 0} P_{\bar{L}} (\hat{\mathbf{p}} \cdot \hat{\mathbf{q}}) \left< p q \alpha | \overline{V}_{123} | p q \alpha' \right> \nonumber \\
&& \times \left\{
\begin{array}{ccc}
 L & L' & \bar{L} \\
 l' & l & \mathcal{L} 
\end{array}
\right\}
\left\{
\begin{array}{ccc}
 L & S & J \\
 l & 1/2 & j \\
 \mathcal{L} & \mathcal{S} & \mathcal{J}
\end{array}
\right\} 
\left\{
\begin{array}{ccc}
 L' & S & J' \\
 l' & 1/2 & j' \\
 \mathcal{L} & \mathcal{S} & \mathcal{J}
\end{array}
\right\}
  \;,
\end{eqnarray}
using standard notation for the angular momentum coupling, the Clebsch-Gordan coefficients 
$\mathcal{C}_{l_1 m_2 l_2 m_2}^{l_3 m_3}$, and $\hat{x} \equiv \sqrt{2 x + 1}$.

In Fig.~\ref{fig:Erho} we show the results for the energy per
neutron at four different resolution scales $\lambda$  as a function of neutron
number density $n$. This range of resolution scales has also been used in 
in previous studies of neutron matter and nuclear matter~\cite{Hebeler_PNM, Hebeler_SNM}
based on $\vlowk$-evolved NN interactions. The present 
calculations use the N$^3$LO NN potential ($\Lambda = 500$ MeV) of 
Ref.~\cite{N3LO} plus the consistent 3NF at N$^2$LO with the couplings 
$c_1 = - 0.81\,\rm{GeV}^{-1}$ and $c_3 = - 3.2\,\rm{GeV}^{-1}$ 
at the initial resolution scale $\lambda = \infty$ $(s=0)$. The other 
components of the full N$^2$LO 3NF give
no contributions in neutron matter~\cite{Hebeler_PNM}. 
We calculate the EOS in three ways: NN-only, 3N-induced and 3N-full
(see Ref.~\cite{Jurgenson_PRL} and the caption to Fig.~\ref{fig:Erho}). The NN 
contributions are resummed and the 3NF contributions are
calculated in Hartree-Fock approximation. When induced 3NF are taken into account,
we find a dramatically reduced $\lambda$ dependence over the
entire density range compared to including only NN forces. 
When initial 3NF are also included the spread of the results increases, but remains still
significantly smaller than the spread of the NN-only results.
The 3N-full energy per neutron for $\lambda=2.0\fmi$ at saturation
density is about 0.5--1.5\,MeV higher than found in calculations
with the same initial NN interaction but with
the unevolved N$^2$LO 3NF included with a cutoff ranging from 
2.0--2.5$\fmi$~\cite{Hebeler_PNM}.  More detailed comparisons will be made in
a future publication.

In Fig.~\ref{fig:Elambda} we show the energy per neutron at
nuclear saturation density for two different NN interactions as a function of
the resolution scale. The left panel uses the NN and 3N interactions as in
Fig.~\ref{fig:Erho} and  the right panel uses the N$^2$LO potential
($\Lambda/\tilde{\Lambda} = 450/500$ MeV)  of Ref.~\cite{Progpart_Epelbaum} plus
the consistent 3NF with the low-energy constants  $c_1 = - 0.81\,\rm{GeV}^{-1}$
and $c_3 = - 3.4\,\rm{GeV}^{-1}$. The bands in both panels at the right side
indicate the size of unevolved second-order contributions containing NN and 3N
forces; these diagrams can be evaluated at the initial RG scale using the
framework of Ref.~\cite{Hebeler_PNM}.
The band is smaller in the right panel due to the smaller initial cutoff of the NN potential.

The variation of the energy in the 3N-full case is always within the width of the band. 
This suggests that the inclusion at finite $\lambda$ of neglected
second-order diagrams with 3NF, which are attractive and should
decrease in magnitude with decreasing $\lambda$~\cite{Hebeler_PNM}, 
may systematically
reduce the observed variation with $\lambda$ above $2.0\fmi$. 
The small increase in energy below $\lambda = 2.0\fmi$ for both
3N-full and 3N-induced curves might be attributed
to induced four-body forces, but it would be premature to draw
quantitative conclusions. 

The inclusion of induced 3NF contributions greatly reduces the resolution-scale
dependence of the EOS  even at the present truncation of the many-body expansion.
In the left/right panel of Fig.~\ref{fig:Elambda} we find a maximal  energy
variation of about 390/470 keV for the 3N-induced calculations and 650/450 keV
for the 3N-full calculations at saturation density. In comparison, we find a total variation 
of about 3.6/4.2 MeV when 3NF are completely neglected. Furthermore, Fig.~\ref{fig:Elambda} also 
demonstrates the increased perturbativeness of the many-body expansion. The solid lines show results using the NN ladder 
sum while the dashed lines are use diagrams up to second order. For the NN interaction at 
N$^2$LO (right panel), diagrams beyond second order in the particle-particle
channel give only very small contributions.

\begin{figure}[b!]
\includegraphics[scale=0.45,clip=]{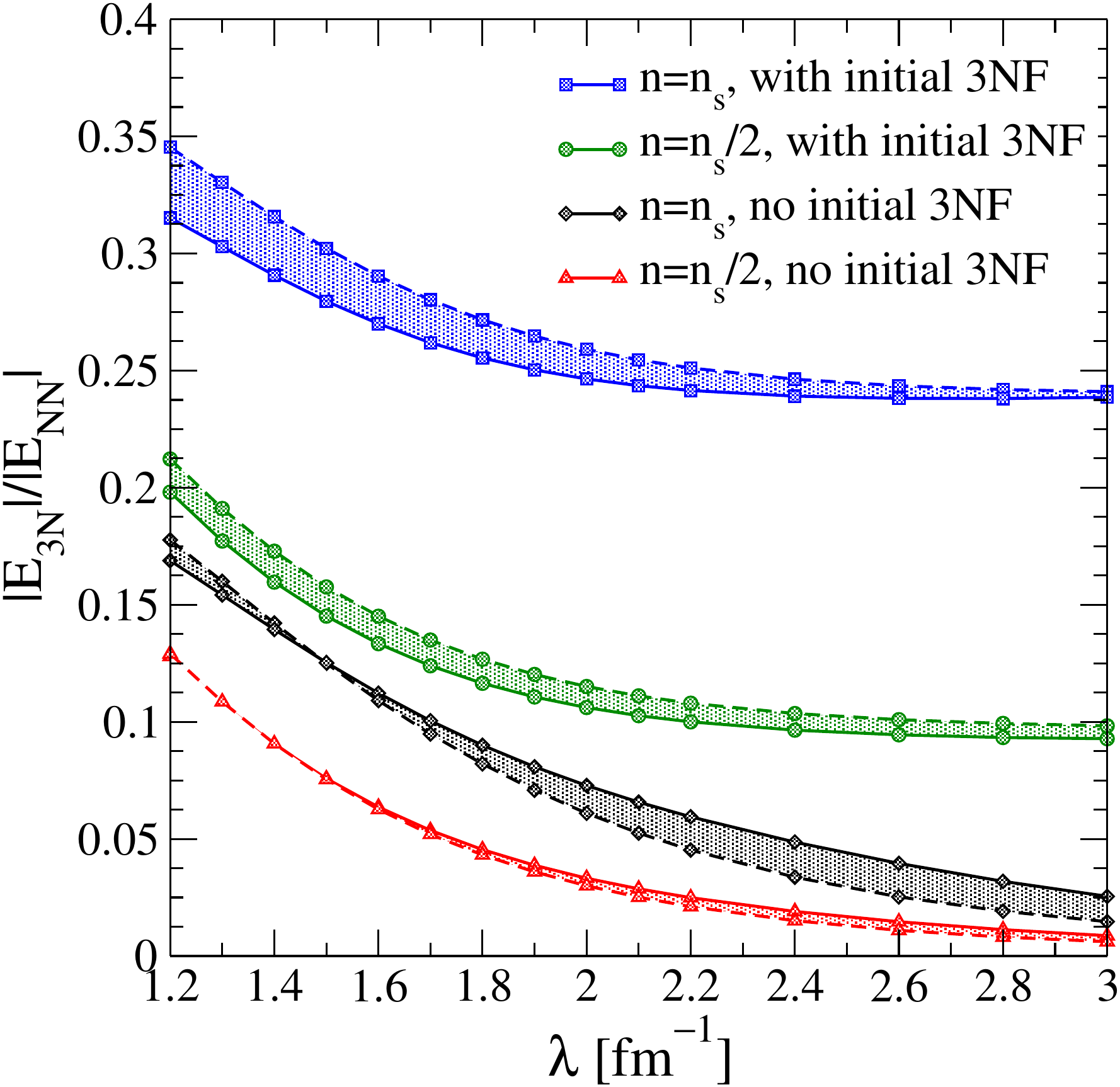}
\caption{(Color online) Scaling of the contributions from NN and 3N forces as a function of $\lambda$. The dashed lines show the ratio of these 
two contributions based on the 500\,MeV N$^3$LO NN potential~\cite{N3LO} and the dashed lines based on the 450/500 MeV N$^2$LO potential~\cite{Progpart_Epelbaum} 
(see also Fig.~\ref{fig:Elambda}).}
\label{fig:scaling}
\end{figure}

\begin{figure*}[t!]
\includegraphics[width=0.95\textwidth,clip=]{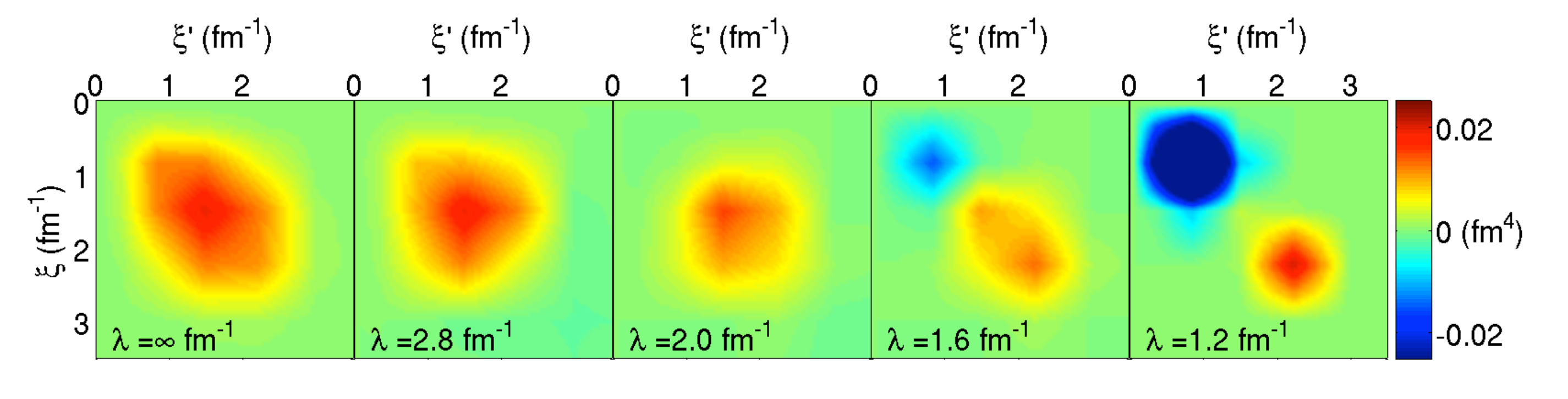}
\caption{(Color online) Matrix elements of the evolved 3N potential $\left< \xi \alpha = 1 | \overline{V}_{123} | \xi' \alpha' = 1 \right>$ (see main text)
for $\mathcal{J}=1/2$, $\mathcal{T}=3/2$ and positive total parity $\pi_3$ at the hyperangle $\theta = \pi/8$ (bottom). The interactions are the same as in Fig.~\ref{fig:Erho}.}
\label{fig:matrix_elements}
\end{figure*}

The net relative growth of two- and three-body contributions to the energy
at two densities are shown in Fig.~\ref{fig:scaling}, with and without
initial 3NF. The solid lines are for the N$^3$LO 500 MeV NN
potential while the dashed lines are for the N$^2$LO 450/500 MeV potential. 
The size of the initial 3NF sets the scale of a natural ratio
at that density. It is evident that the change of the net ratio
with $\lambda$ remains natural even down to the smallest values.
There is no obvious trend with density; such a trend may be obscured by
cancellations among contributions to the net energies.

We can also examine the evolution of the 3NF by looking at slices
of the matrix elements. To do this we introduce the hyperradius $\xi^2 = p^2 + 3/4 q^2$ and
the hyperangle $\tan \theta = 2p/(\sqrt{3} q)$ and visualize the matrix elements as a function of $\xi$ at a fixed hyperangle. 
A representative example is shown in Fig.~\ref{fig:matrix_elements} for $\theta = \pi/8$ for the dominant 
partial wave with $\alpha = \alpha'$ and $L=J=S=l=0$, $T=1$, $j=1/2$, $\mathcal{J}=1/2$ and $\pi_3 = 1$. In this case
we observe softening from $\lambda = \infty$ to $\lambda = 2.0$ fm$^{-1}$. 
Further evolution causes a strongly attractive part to appear at small
momenta; however, its impact will be mitigated by phase space factors.
This renormalization merits further study.

In this paper we have presented the first neutron matter calculations
based on a fully consistent RG evolution of two- and three-body
chiral EFT interactions. Including induced 3NF greatly reduces the resolution scale 
dependence of the neutron EOS compared to NN-only calculations, with the 
residual dependence comparable to the expected magnitude of omitted many-body
corrections. Thus we are not able to make a definitive statement about
the size of induced four-body contributions, but there are no indications
of unnatural growth. Future calculations will include both neutron and nuclear matter with
higher-order (beyond Hartree-Fock) diagrams including 3NF, which
will allow a more complete assessment of higher-body contributions.
We will also calculate error bands based on uncertainties in the
3NF input. In the future it will be also straightforward to include contributions from 3NF 
at N$^3$LO~\cite{Bernard_N3LO1, Bernard_N3LO2} once initial matrix elements in 
partial wave representation are available. This work is currently in progress.

\medskip

\begin{acknowledgments}
We thank H.\ Hergert, for helpful discussions and H.~Hergert, T.~Kr\"uger, 
A.~Schwenk, V.~Soma, and I.~Tews for
useful comments on the manuscript. We are grateful to J.\ Golak, R.\ Skibinski and K.\ Topolnicki for 
providing us a code for generating the antisymmetrized matrix elements 
of the initial 3N forces. This work was supported in part by the National Science Foundation 
under Grant No.~PHY--1002478, the U.S.\ Department of Energy under Grant No.~DE-SC0008533 (SciDAC-3/NUCLEI project), and 
an award of computational resources from the Ohio Supercomputer Center.
\end{acknowledgments}

\end{document}